\documentclass[sigconf]{acmart}

\setcopyright{acmcopyright}
\copyrightyear{2023}
\acmYear{2023}
\acmDOI{XXXXXXX.XXXXXXX}

\acmConference[KDD '23]{KDD '23}{August 06-10, 2023}{Long Beach, CA, USA}
\acmPrice{15.00}
\acmISBN{978-1-4503-XXXX-X/23/08}

\begin{document}

\title{Rethinking Financial Service Promotion With Hybrid Recommender Systems at PicPay}

\author{Gabriel Mendonça}
\email{gabriel@land.ufrj.br}
\orcid{0000-0002-2294-0103}
\affiliation{%
  \institution{Federal University of Rio de Janeiro}
  \city{Rio de Janeiro}
  \country{Brazil}
}
\affiliation{%
  \institution{PicPay}
  \city{São Paulo}
  \country{Brazil}
}

\author{Matheus Santos}
\email{matheus.salmeida@picpay.com}
\affiliation{%
  \institution{PicPay}
  \city{São Paulo}
  \country{Brazil}
}

\author{André Gonçalves}
\email{andre.sgoncalves@picpay.com}
\orcid{0009-0005-0550-4966}
\affiliation{%
  \institution{PicPay}
  \city{São Paulo}
  \country{Brazil}
}

\author{Yan Almeida}
\email{yan.almeida@picpay.com}
\affiliation{%
  \institution{PicPay}
  \city{São Paulo}
  \country{Brazil}
}

\renewcommand{\shortauthors}{Mendonça, et al.}

\begin{abstract}
The fintech PicPay offers a wide range of financial services to its 30 million monthly active users, with more than 50 thousand items recommended in the PicPay mobile app. In this scenario, promoting specific items that are strategic to the company can be very challenging. In this work, we present a Switching Hybrid Recommender System that combines two algorithms to effectively promote items without negatively impacting the user's experience. The results of our A/B tests show an uplift of up to 3.2\% when compared to a default recommendation strategy.
\end{abstract}




\maketitle

\section{Introduction}


With 30 million monthly active users, PicPay~\cite{picpay} is one of the largest fintechs in Brazil. The company offers a large catalog of products and services through its mobile app, totaling more than 50 thousand distinct items. Similarly to super-apps like WeChat~\cite{montag2018multipurpose,wechat} in China and Gojek~\cite{gojek} in Indonesia, the PicPay app addresses multiple customer necessities, such as tax payments, loans, gift cards, coupons, instant money transfers, investments, and cryptocurrency transactions.

\begin{figure}[ht]
    \centering
    \includegraphics[width=0.65\linewidth]{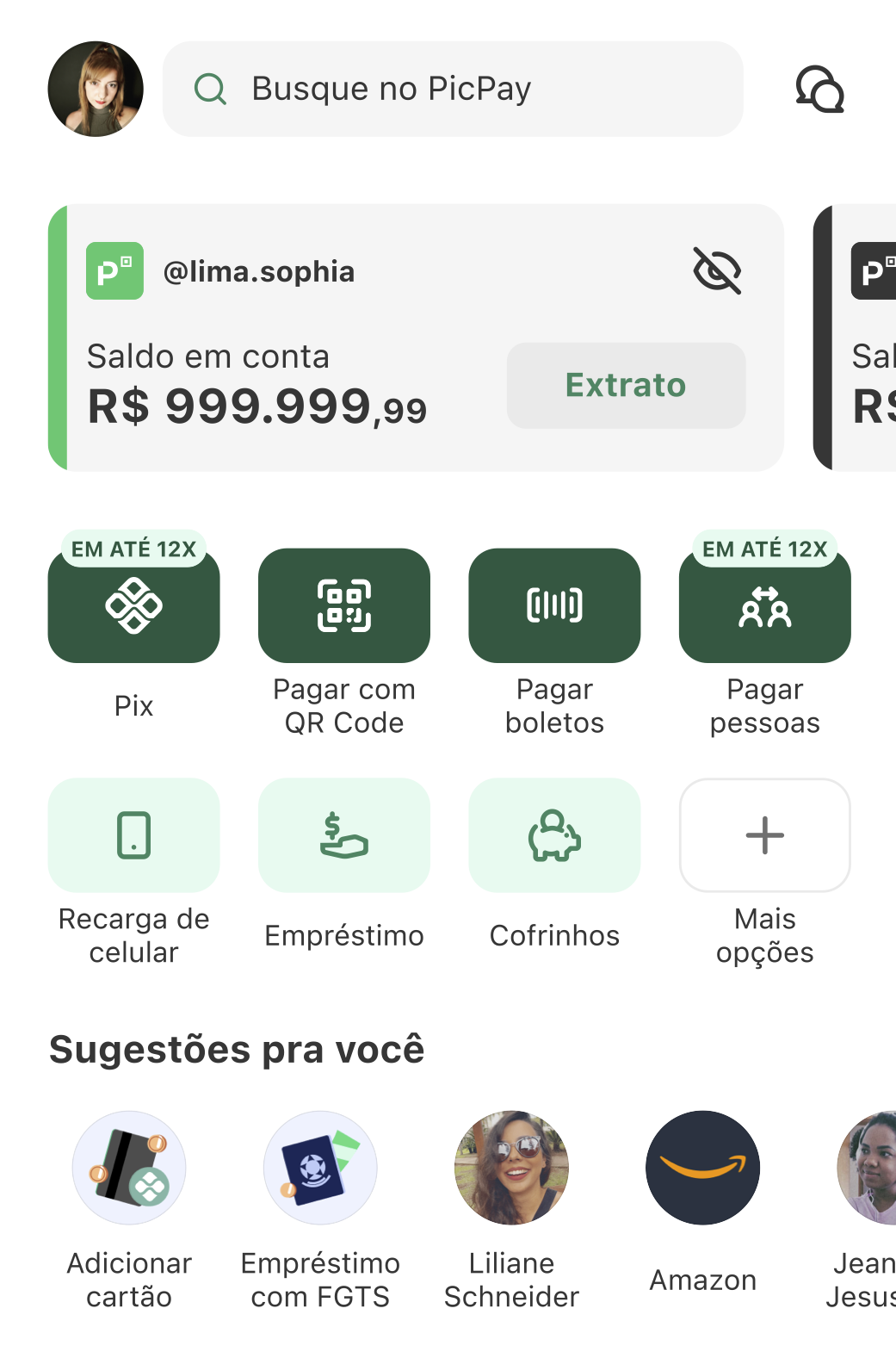}
    \caption{Home screen of the PicPay app.}
    \label{fig:home}
\end{figure}

Such an extensive and diverse catalog favors the adoption of Recommender Systems. Along with some shortcut buttons and basic account information, the PicPay app's home screen contains a unique UI element that is customized for each user. The ``\textit{Suggestions for You}'' component consists of a list of 15 items that are recommended based on multiple consumer features, such as credit profile, purchase history, and behavior profile. 
Figure~\ref{fig:home} shows the UI of the mobile app, with the \textit{Suggestions for You} (in Portuguese, ``\textit{Sugestões pra você}'') component displayed at the bottom. 

The \textit{Suggestions for You} component is often used as an instrument in awareness campaigns to highlight specific products and services that are strategic for the company (e.g., encouraging upselling or cross-selling).
With limited available space, we need to carefully choose the items that will be shown to each user to avoid the pitfalls of low recommendation diversity~\cite{10.1145/1250910.1250939} and excessive advertising~\cite{iyer2020impulse}. As an additional challenge, we need to be able to handle cold-start scenarios when promoting new products.

In this paper, we present the \textit{hybrid recommender system}~\cite{burke2002hybrid} used by PicPay to efficiently recommend strategic items to millions of users every day.

\section{Related work}

Falk and Karako~\cite{10.1145/3523227.3547393} discuss the problem of choosing between multiple recommendation models for Shopify merchants depending on how much historical data are available for each store. Unlike Shopify, items recommended at PicPay are not equally important. For example, we may want to promote products that are new, generate more revenue, or increase user engagement. Therefore, our recommendation models must be tailored according to our business strategies.

Other authors~\cite{10.1145/3209978.3209993,10.1145/3460231.3474621} propose to weight prediction errors in Recommender Systems based on expected revenue. At PicPay, we cannot establish an item priority based on a simple metric like product price or profit margin since recommending ``enablers'' (e.g., adding a credit card to our digital wallet) can lead to future rewards that are not easily measured.

\section{\textit{Suggestions for You}}

One of the main elements on the PicPay app's home screen, the \textit{Suggestions for You} component displays a list of 15 items recommended by a hybrid recommender system~\cite{burke2002hybrid}. Each item acts as a call to action (CTA), with alternatives such as transfering money to a friend, buying credits for a prepaid SIM card or applying for a personal loan. Due to their strategic value to the company, a small subset of items is regarded as \textbf{\textit{promoted items}}. \textit{Promoted items} may be associated with new products, profitable financial services, or ``enablers'' (e.g., adding a credit card to your digital wallet, consenting to an Open Banking API). The remaining recommended items in the \textit{Suggestions for You} component are regarded as \textbf{\textit{regular items}}.

\begin{figure}[ht]
    \centering
    \includegraphics[width=0.8\linewidth]{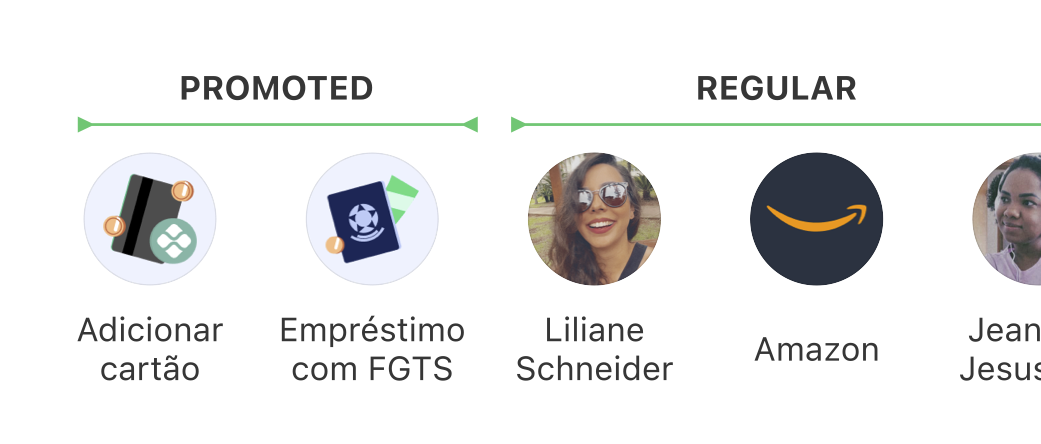}
    \caption{\textit{Suggestions for You} UI component}
    \label{fig:sugestoes}
\end{figure}

The size of the phone screen typically restricts the exhibition to four items (the fifth item is cropped). Viewing the latter items in the list requires the user to perform a swipe gesture. Figure~\ref{fig:sugestoes} shows a sample recommendation. On the left, two \textit{promoted items} are presented, followed by three \textit{regular items}. As we can see, the \textit{promoted items} in the \textit{Suggestions for You} component are always displayed \textit{before} the \textit{regular items}.

We confirm the hypothesis of increased exposure at the first positions of the component through an online experiment. Without altering \textit{what} was recommended to each user, we \textit{shuffled} each list of recommendations, displaying the items in \textit{random order}. As a consequence, users who participated in the experiment were able to observe some \textit{regular items} taking precedence over \textit{promoted items}. With all \textit{promoted} and \textit{regular} items having the same probability of being displayed in each position, we estimate the \textit{intrinsic} click-through rate (CTR) of each of the 15 positions.

\begin{figure}[ht]
    \centering
    \includegraphics[width=1\linewidth]{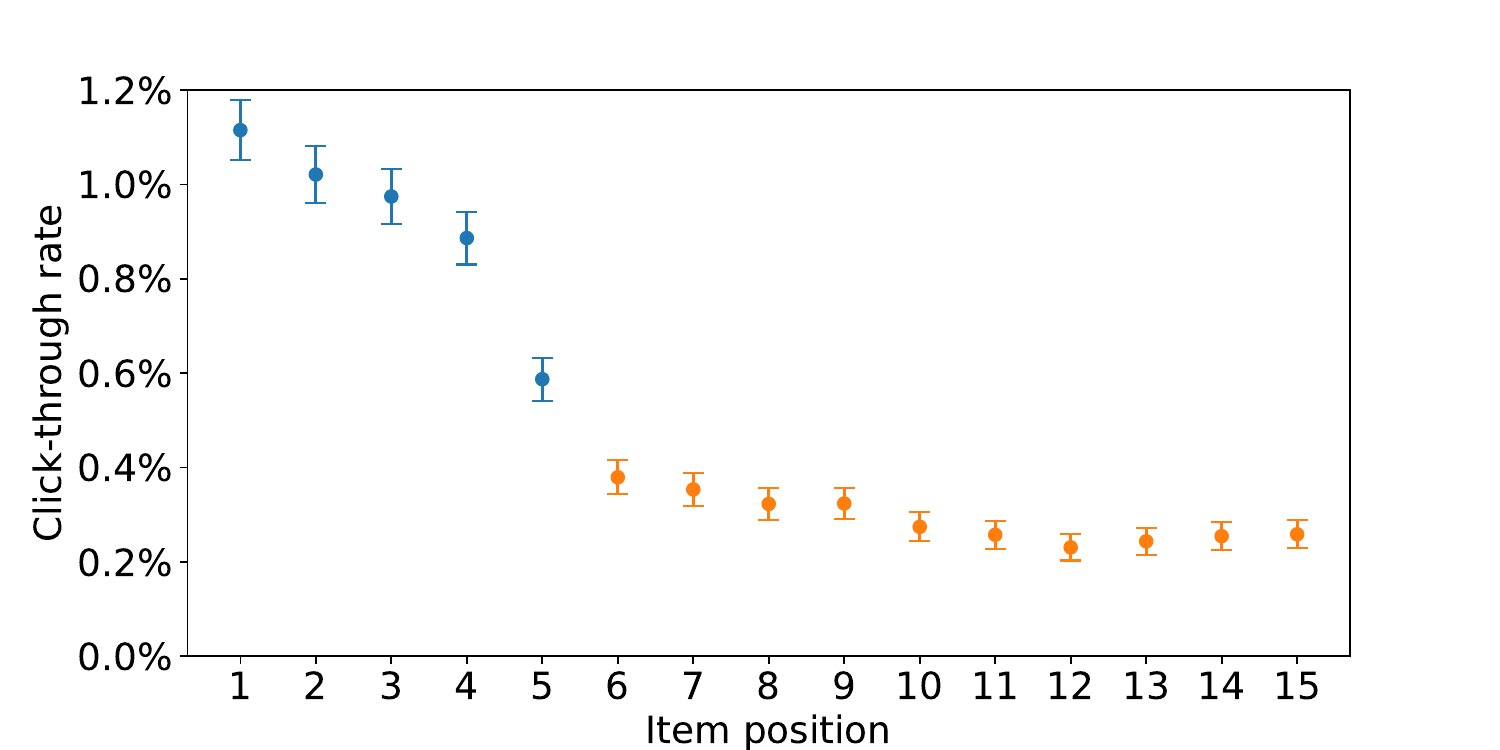}
    \caption{Probability of clicking for each position of the \textit{Suggestions for you} component.}
    \label{fig:click_model}
\end{figure}

During a period of 10 days, we displayed the \textit{shuffled} version of the \textit{Suggestions for You} component for a random sample of 28,235 users. Figure~\ref{fig:click_model} shows the results of the experiment with a 95\% Confidence Interval. We can see that positions immediately visible when the app is opened (in blue) tend to receive more clicks than those that require a swipe gesture from the user to be seen (in orange). We cannot clearly identify whether users follow a Cascade Model~\cite{craswell2008experimental}, scanning recommendations from left to right. However, we can see that the first four positions can receive, on average, up to five times more clicks than the latter positions.

By revealing higher click-through rates at the first positions of the component, the experiment confirmed the risks of recommending \textit{promoted items} that are not relevant to the user. Consequently, \textit{promoted items} should be carefully recommended due to their priority over \textit{regular items}.

\section{Recommending \textit{promoted items}}

Conventional collaborative filtering and content-based approaches do not bode well for the recommendation of financial services~\cite{Felfernig2015}. Contrary to content such as videos, books, and news, financial services typically have scarse metadata and user feedback. For example, services like loans are not contracted very often. In this context, knowledge-based approaches, which employ deep domain-specific knowledge, are preferable~\cite{Felfernig2015}.

In our mobile app, we recommend \textit{promoted items} in the \textit{Suggestions for You} component using a constraint-based approach. PicPay's domain experts are responsible for building a \textit{recommender knowledge base}~\cite{Felfernig2015}, that is, a set of variables and constraints that explicitly represent the criteria needed for the recommendation of a \textit{promoted item} to a given user. In this process, multiple information sources are taken into account. The offer of credit cards, for example, depends on the user's credit profile, their risk analysis, and other features\footnote{Further details are ommited due to the strategic value of the information.}.

\begin{figure}[ht]
    \centering
    \includegraphics[width=0.8\linewidth]{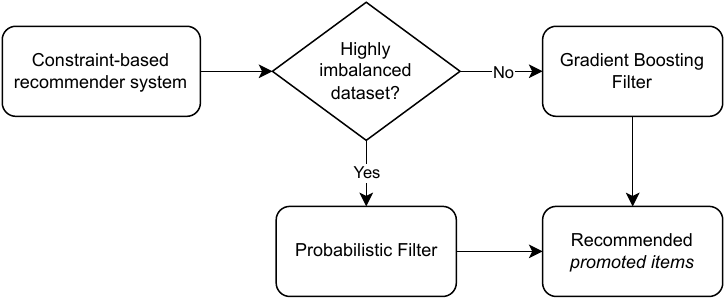}
    \caption{Switching Hybrid Recommender System for promoted items.}
    \label{fig:ensemble}
\end{figure}

For some users, there may be multiple \textit{promoted items} that satisfy the constraints. Therefore, we need to further refine the recommendations to increase the number of relevant items in the first (and most noticeable) positions, as well as achieve a good balance between \textit{regular} and \textit{promoted} items. We apply two post-filtering methods that operate in tandem, composing the \textit{Switching Hybrid Recommender System} illustrated in Figure~\ref{fig:ensemble}. Using this architecture, we benefit from the strengths of two different techniques, switching between two recommendation techniques according to a chosen criterion~\cite{burke2002hybrid}.

The first method, the \textit{Probabilistic Filter} algorithm, can be applied to any item and does not require training. Therefore, it is better suited for items with a very low number of conversions (e.g., new products). The \textit{Gradient Boosting Filter} uses an ML classifier to improve the relevance of recommended items and thus may give varying results depending on the available data. As shown in Figure~\ref{fig:ensemble}, the switch criterion is based on the fraction of positive samples available for each \textit{promoted item}. We use the \textit{Probabilistic Filter} when the dataset is highly imbalanced, that is, less than 1\% of the samples represent a conversion.

In the following, we describe the two methods in detail.

\subsection{Probabilistic Filter}

Ideally, the \textit{promoted items} recommended by our knowledge-based system should be filtered according to the user's interest. However, when promoting new products or services in the \textit{Suggestions for You} component, it is impractical to leverage past user interactions to predict if an item is useful to a given customer. In this situation, we need a filtering algorithm that: (1) does not require training with historical data; (2) is efficient, since the app receives millions of accesses every day; and (3) avoids \textit{starvation}, i.e., given a threshold $N_s$, an item should be seen by the customer with high probability after opening the app $N_s$ times. Although a round-robin algorithm would lead to a fair allocation~\cite{103550}, it would not be very efficient because we would need to keep track of the \textit{promoted items} viewed by each user.

Inspired by random scheduling policies in queue systems~\cite{conway1967scheduling}, we propose a \textit{Probabilistic Filter} algorithm. When a customer opens the application, each recommended \textit{promoted item} is sampled independently with \textbf{\textit{display probability} $p$}. This parameter allows for the adjustment of the (expected) number of \textit{promoted items} seen by the user. While a low value of $p$ leads to more space available for \textit{regular items}, a high value $p$ decreases the odds of starvation of \textit{promoted items}.

As an additional benefit, reducing the frequency of exhibition can have positive impacts on the item, since a product exposed in an exaggerated way may arouse less user interest compared to a product with limited exposure~\cite{iyer2020impulse}.

Let $N$ be the number of times the app has to be opened until the item appears for the first time for a given user. Since the item has a probability $p$ of appearing and each attempt is independent, $N$ will follow a geometric distribution.
Then we define the \textbf{\textit{probability of starvation}} $\alpha$ as

\begin{displaymath}
    \alpha \triangleq Pr(N > N_s) = (1-p)^{N_s},
\end{displaymath}
where $N_s \in \mathbb{N}$ is the \textbf{\textit{visualization threshold}} parameter. The parameter $\alpha$ can be interpreted as the probability that a consumer will not see a specific \textit{promoted item} after opening the app $N_s$ times. Solving for $p$, we have

\begin{displaymath}
    p = 1 - \alpha^{1 / N_s}
\end{displaymath}

As a result, we can choose the display probability $p$ of each \textit{promoted item} as a function of the desired probability of starvation $\alpha$ and visualization threshold $N_s$.

Since items are sampled independently, the number of items that are displayed will follow a Poisson binomial distribution with parameters $p_1, p_2, \ldots, p_m$, where $p_i$ is the display probability of the $i$th item. As a consequence, the expected number of \textit{promoted items} will be given by $\sum_{i=1}^{m} p_i$. In the special case where all display probabilities are the same, the number of \textit{promoted items} will follow a binomial distribution.



\subsection{Gradient Boosting Filter}

When a \textit{promoted item} has a record of user feedback (e.g., purchases, conversions, or clicks), we can develop an ML classifier to predict future user interest and use the classifier's output to filter items with low probability of conversion.

We train one LightGBM classifier~\cite{NIPS2017_6449f44a} per \textit{promoted item}. Each sample in the dataset corresponds to a customer, and the label indicates whether the user converted at least once during a 90-day period. We use a total of 25 features, including: credit score, device OS, age, income, address region, and RFM (Recency, Frequency, Monetary value) features. Since gradient-boosted trees models have problems with calibration~\cite{10.1145/1102351.1102430}, we fit a nonparametric isotonic regressor~\cite{10.1145/775047.775151} to adjust the probability values output by the model.

Given a classifier trained to predict conversions for a \textit{promoted item} \textit{i}, the \textit{Gradient Boosting Filter} displays the item \textit{iff} the classifier's output is above a threshold $T_i$. We choose each threshold value based on the analysis of the cumulative gains curve, which is often used in direct marketing campaigns~\cite{10.5555/3000292.3000304}. By evaluating the classifier at different threshold values, the cumulative gains curve provides an estimate of the number of users we need to impact in order to reach a certain fraction of the users interested in the product.



\section{Results}

We evaluate our algorithms through two A/B tests. Both experiments were carried out for 14 days and the main metric was the \textit{Suggestions for You} component's overall conversion rate. As guardrail metrics, we also measured the individual conversion rate of each \textit{promoted item}.

\paragraph{Probabilistic Filter}
In the first experiment, we selected a random sample of 1.5M users, who were divided into two groups. Users in the control group received the direct output of the constraint-based recommender system (see Figure~\ref{fig:ensemble}), while users in the treatment group received the recommended \textit{promoted items} with the \textit{Probabilistic Filter}. For the four items considered, we set $N_s=5$ and $\alpha=0.1$, leading to $p=0.37$. Given the previous behavior, we expected users to see each item in up to 7 days with a probability of 0.9. On average, each user in the treatment group saw $4 \cdot p = 1.48$ \textit{promoted items} when opening the app.

Comparing the average conversion rate of the two groups, we observed a statistically significant uplift of 3.2\%  ($p < 0.0001$). Furthermore, there was no statistically significant evidence of a decrease in the conversion rate of the \textit{promoted items}. The results suggest that it is possible to improve the customer experience by reducing the frequency of exhibition of \textit{promoted items} without negative impacts on the conversion rate of \textit{promoted items}.

\paragraph{Gradient Boosting Filter}
A total of 4.4M users participated in the second A/B test, considering 3 \textit{promoted items}. For users in the control group, the \textit{promoted items} were recommended using the \textit{Probabilistic Filter}. The users in the treatment group received recommendations using the \textit{Gradient Boosting Filter}. When selecting the \textit{Gradient Boosting Filter}'s threshold value, we considered the classifier's recall, the expected fraction of customers impacted by the recommendation, and the relevance of the \textit{promoted item} to the business. We show the chosen threshold values in Table~\ref{tab:gbthresholds}.

\begin{table}
  \caption{Threshold values for the \textit{Gradient Boosting Filter}.}
  \label{tab:gbthresholds}
  \begin{tabular}{cccc}
  \toprule
  Promoted item & Threshold & Recall & Impacted users \\
  \midrule
  ``Credit Card''   & 0.10 & 0.88 & 20.6\%\\
  ``Open Banking''  & 0.10 & 0.67 & 37.8\%\\
  ``Personal Loan'' & 0.04 & 0.55 & 11.4\%\\
  \bottomrule
\end{tabular}
\end{table}

As in the first experiment, the results were positive, with an increase of 1.1\% in the conversion rate ($p = 0.02$). We can see that, when data from previous conversions are available, the \textit{Gradient Boosting Filter} can further improve the effectiveness of the recommendation of \textit{promoted items}.



\section{Conclusion}

Every day, PicPay serves a vast catalog of financial services to millions of users through its Recommender Systems. In this context, promoting strategic products and services is not easy, as these promoted items may not be as relevant to the consumer as the ones commonly consumed.
In this paper, we present our solution to promote financial services using a Switching Hybrid Recommender System. Our system employs two algorithms: a \textit{Probabilistic Filter} that is very scalable and can be easily parameterized; and a \textit{Gradient Boosting Filter} that uses an ML classifier to predict user interest. The results of our A/B tests show that our algorithms can increase conversion rates by at least 1.1\%.

\begin{acks}
We would like to acknowledge our team members who contributed to this project: Francielle Fernandes, Gustavo H. A. Santos, Lucas Moraes, Matheus Barbieri, and Sophia Lima.
\end{acks}

\bibliographystyle{ACM-Reference-Format}
\bibliography{article}

\end{document}